\documentclass[twocolumn,showpacs,preprintnumbers,amsmath,amssymb]{revtex4}
\usepackage{tabularx,graphicx}\begin{document}
\newcommand{\beq}{\begin{equation}}
\newcommand{\eeq}{\end{equation}}
\newcommand{\beqn}{\begin{eqnarray}}
\newcommand{\eeqn}{\end{eqnarray}}
\newcommand{\bmath}{\begin{subequations}}
\newcommand{\emath}{\end{subequations}}
\title{Correcting 100 years of misunderstanding: electric fields in superconductors, hole superconductivity,  and the Meissner effect}
\author{J. E. Hirsch }
\address{Department of Physics, University of California, San Diego,
La Jolla, CA 92093-0319}

\begin{abstract} 
From the outset of superconductivity research it was assumed that no electrostatic fields could exist inside superconductors, and this assumption was incorporated into conventional London electrodynamics. Yet  the London brothers themselves initially (in 1935) had proposed an electrodynamic theory of superconductors that allowed for static electric fields in their interior, which they unfortunately discarded a year later. I argue that the Meissner effect in superconductors necessitates  the existence of an electrostatic field in their interior, originating in the expulsion of negative charge from the interior to the surface when a metal becomes superconducting.  The theory of hole superconductivity predicts this physics, and associated with it a macroscopic spin current in the ground state of superconductors (``Spin Meissner effect''), qualitatively different from what is predicted by conventional BCS-London theory. A new London-like electrodynamic description of superconductors is proposed to describe this physics. Within this theory superconductivity is driven by lowering of quantum kinetic energy, the fact that the Coulomb repulsion strongly depends on the character of the charge carriers, namely whether electron- or hole-like, and the spin-orbit interaction. The electron-phonon interaction does not play a significant role, yet the existence of an isotope effect in many superconductors is easily understood.  In the strong coupling regime the theory appears to favor local charge inhomogeneity. The theory is proposed to apply to all superconducting materials, from the elements to the high $T_c$ cuprates and pnictides, is highly falsifiable, and explains a wide variety of experimental observations.
    \end{abstract}
\maketitle 

\section{introduction}

The eminent experimental physicist K. Mendelssohn wrote in 1966\cite{mend}  {\it ``To the layman it may come as a disappointment that the explanation of such a striking
phenomenon as superconductivity should, on the atomistic scale, have been revealed as nothing more exciting than a footling small
interaction between electrons and lattice vibrations. This feeling was shared by many physicists who had hoped that superconductivity might reveal some new
fundamental principle of nature''}, undoubtedly including himself in that physicists' group. The work discussed in this paper suggests that Mendelssohn's disappointment may
have been premature.

In the conventional theory of superconductivity the speed of light $c$ plays no role, as pointed out  by Alexandrov\cite{alexandrov}. Whether it is
$300,000 km/s$ or $3mm/s$ would not change the magnitude of $T_c=4.153^oK$ for mercury, as discovered by Kammerlingh Onnes $100$ years ago\cite{kamm}, 
nor the isotope coefficient  $\alpha=0.5$ of mercury `discovered' theoretically by Fr\"{o}hlich $60$ years ago\cite{fro}. 

However, E.U. Condon wrote a paper in 1949 (just before the discovery of the isotope effect)  entitled ``Superconductivity and the Bohr magneton''\cite{condon}. He pointed out 
in that paper that the maximum magnetization measured in superconducting cylinders of various materials appears to have an empirical relationship with the magnetization that would result from the conduction electrons
intrinsic magnetic moment due to spin, all alligned. This suggests an intrinsic relationship between the Bohr magneton, which contains the speed of light,
and the superconducting state. 

Indeed, consider the following expression for the lower critical field of type II superconductors:
\beq
H_{c1}=-\frac{ \hbar c}{4e\lambda_L^2}
\eeq
with $e$ the electron charge and the London penetration depth $\lambda_L$ given by the usual expression\cite{tinkham}
\beq
\frac{1}{\lambda_L^2}=\frac{4\pi n_s e^2}{m_e c^2}
\eeq
with $n_s$ the number of superconducting electrons of mass $m_e$ per unit volume. Eq. (1) follows if the flux through an area of radius $2\lambda_L$ is
the flux quantum $\phi_0=hc/2e$. (The usual expression for $H_{c1}$ derived from Ginzburg Landau theory\cite{tinkham} has an extra logarithmic factor of order $1$).
The magnetization of a cylindrical sample in the presence of an external magnetic field of magnitude just below $H_{c1}$ is
\beq
M_c=\frac{H_{c1}}{4\pi}=\frac{1}{2} n_s \mu_B
\eeq
with $\mu_B=e \hbar/2m_e c$ the intrinsic magnetic moment of the electron (Eq. (2) was used in obtaining Eq. (3)). $(1/2)n_s$ is the number of superconducting electrons of each spin.

Thus, we can think of the superconductor in the absence of applied field as having magnetization zero arising from the cancellation of $+M_c$ and $-M_c$ originating
in the intrinsic magnetic moments of the conduction electrons with spin down and up, as shown schematically in Fig. 1(a). When an external magnetic field
is applied, the superconductor develops a magnetization due to the induction of surface currents, and when that magnetization reaches the value $M_c$ the
superconducting state is destroyed. This suggests  that the superconducting state $``knows''$ about $H_{c1}$ and $M_c$ before the magnetic field is applied. 
If the superconducting state knows about $M_c$, it knows about the magnitude of the magnetic moment of the electron and the speed of light.
Instead, in the conventional theory of superconductivity, the spin of the electron is important only insofar as it can point in two directions and allow for
Cooper pairing $(k\uparrow,-k\downarrow)$, but the associated magnetic moment plays absolutely no role and may as well be zero
(as would be the case for $c\rightarrow \infty$).

   \begin{figure}
 \resizebox{7.5cm}{!}{\includegraphics[width=9cm]{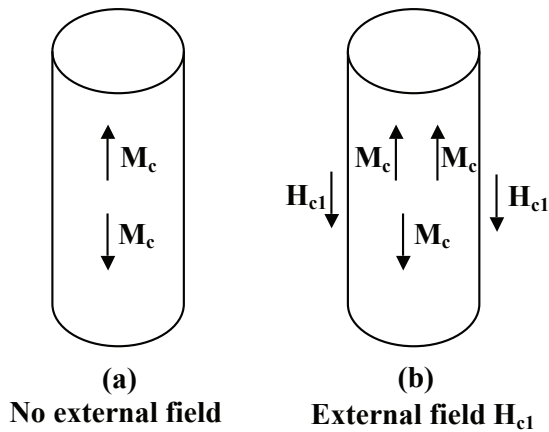}}
 \caption {Superconducting cylinder in the absence of applied external fields. There is no net magnetization, which can be thought of as the superposition
 of the magnetization $M_c=(n_s/2)\mu_B$ generated by the electron spin magnetic moments pointing in opposite directions. (b) When an external magnetic field
 of magnitude slightly less than $H_{c1}$ is applied, an extra magnetization $M_c$ is generated in one direction, and when the field is increased further superconductivity
 is destroyed.}
 \label{figure2}
 \end{figure}

Condon's paper did not have  great impact: when this paper is published it will double the total citations to it,  according to the
Web of Science database. Presumably this is at least in part because the attention of physicists working on superconductivity 
 shifted to the electron-phonon interaction after $1950$. The theory discussed here suggests, in agreement with Condon, that the intrinsic magnetic moment of
the electron plays a key role in superconductivity.

Another seminal work that was lost in the later developments was the first electrodynamic theory of the London  brothers\cite{london1}. It made the bold proposition that 
the `perfect conductor' equation for the supercurrent $\bold{J}$  resulting from an electric field $\bold{E}$
\beq
\frac{\partial \bold{J}}{\partial t}=\frac{n_se^2}{m_e}\bold{E}
\eeq
was in fact invalid for superconductors, and led to a system of equations predicting that $both$ magnetic and electrostatic fields would be screened
in superconductors over a length $\lambda_L$. Note that this would imply that $c$ enters in the behavior of superconductors in the absence
of applied magnetic fields. However this version of the theory was discarded by the London brothers themselves  in favor of the 
currently accepted version shortly after being proposed, 
partly because of theoretical criticism of it by von Laue\cite{laue} and partly because the results of an experiment performed by H. London shortly thereafter appeared to indicate
that it was invalid\cite{hlondon}. In the conventional London theory\cite{londonbook}  Eq. (4) is valid which implies that no electrostatic fields can exist
in the interior of superconductors.

 \begin{figure}
 \resizebox{8.5cm}{!}{\includegraphics[width=9cm]{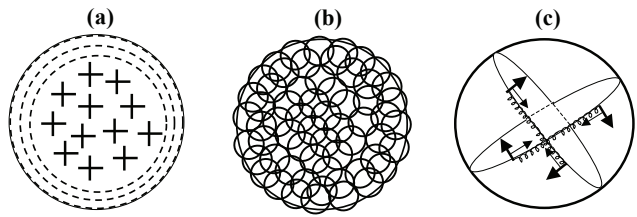}}
 \caption {Illustration of three key aspects of the physics of superconductors proposed here. (a) Superconductors expel negative charge from their
 interior to the region near the surface; (b) Carriers reside in mesoscopic overlapping orbits of radius $2\lambda_L$
 ($\lambda_L$=London penetration depth); (c) A spin current flows near the surface of superconductors (the arrow perpendicular to the orbit denotes the direction
 of the electron magnetic moment). }
 \label{figure2}
 \end{figure}

We have proposed that superconductivity is associated with lowering of the electronic kinetic energy\cite{kinetic1,kinetic2,kinetic3}, expansion of electronic orbits from the
microscopic scale $k_F^{-1}$ ($k_F=$ Fermi wavevector) to the mesoscopic scale $2\lambda_L$\cite{sm} and expulsion of negative charge from the interior of the
superconductor to a surface layer of thickness $\lambda_L$\cite{chargeexp}.  This results in a superconducting ground state where the charge distribution is
macroscopically inhomogeneous and an electric field exists everywhere in the interior of the superconductor (except for a single point) pointing towards the nearest
surface\cite{electrodyn}. It also results in the existence of a macroscopic spin current flowing in the ground state within a London penetration depth of the surface, a kind of
zero-point motion of the superfluid\cite{sm}. This physics is shown schematically in Fig. 2.

The charge electrodynamics of superconductors is described by the four-dimensional equation\cite{electrodyn}
\beq
J-J_0=-\frac{c}{4\pi \lambda_L^2} (A-A_0)
\eeq
with the four-vectors
\bmath
\beq
A=(\bold{A},i\phi) ; J=(\bold{J},ic\rho)
\eeq
\beq
A_0=(0,i\phi_0); J_0=(0,ic  \rho_0)
\eeq
\emath
The potential $\phi_0$ results from the existence of a uniform positive charge density $\rho_0$ in the interior of the superconductor
\beq
\phi_0(\bold{r})=\int_V d^3r' \frac{\rho_0}{|\bold{r}-\bold{r}'|}
\eeq
and the magnitude of $\rho_0$ is fixed by the requirement that the average electric field near the surface is
\beq
E_m=-\frac{\hbar c}{4e\lambda_L^2}
\eeq
i.e. the same as $H_{c1}$, Eq. (1) (in c.g.s. units). For a spherical or cylindrical sample this results in a uniform negative charge density $\rho_-$ 
within a London penetration depth of the surface of magnitude
\beq
\rho_-=-\frac{E_m}{4\pi \lambda_L}
\eeq
and the interior charge density $\rho_0$ is given by
\beq
\rho_0=-\frac{2\lambda_L}{R}\rho_-   ;   \rho_0=-\frac{3\lambda_L}{R} \rho_-
\eeq
for a cylindrical and spherical sample respectively.

The electric potential in the interior of the superconductor satisfies the differential equation
\beq
\phi(\bold{r})=\lambda_L^2\nabla^2\phi(\bold{R})+\phi_0(\bold{r})
\eeq
and can be found analytically for simple geometries (sphere, cylinder, plane)\cite{electrospin} and numerically for other cases, to find the charge distribution and electric field
both inside and outside the superconductor. Boundary conditions assumed are that   $\phi$ and its normal derivative are continuous at the surface of the superconductor,
so we assume that no two-dimensional surface charge exists. For a spherical geometry, infinite cylinder and infinite plane, no electric fields exist outside the
superconductor. For other geometries however electric field lines do exist outside the superconductor, implying that the surface is not equipotential.

  \begin{figure}
 \resizebox{7.5cm}{!}{\includegraphics[width=9cm]{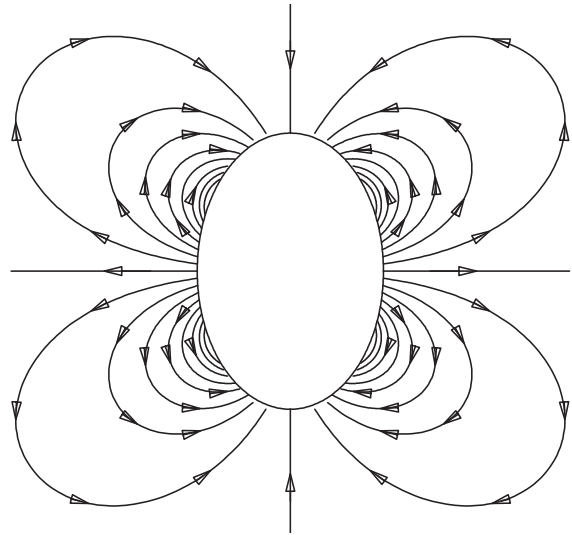}}
 \caption {Electric field lines in the exterior of a sample of ellipsoidal shape obtained from numerical solution of Eq. (11). }
 \label{figure2}
 \end{figure}

   \begin{figure}
 \resizebox{8.5cm}{!}{\includegraphics[width=9cm]{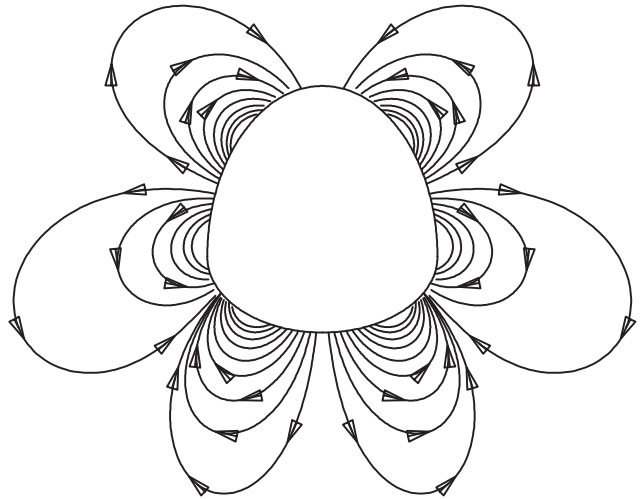}}
 \caption {Electric field lines in the exterior of a sample of egg-like  shape obtained from numerical solution of Eq. (11). }
 \label{figure2}
 \end{figure}
 
Figure 3 shows a representative example of electric field lines obtained by numerical solution of the differential equation for the case of an
ellipsoid\cite{ellipsoid}. The field is quadrupolar in nature, and it can be seen that electric field lines come out of the region of low surface curvature
and come in in the region of high surface curvature. This is understood qualitatively as follows: electrons in the spin current near the surface can move
faster when the surface has less curvature, just like a racing car, hence they have higher kinetic energy and consequently lower potential energy, 
than slower-moving electrons in regions of higher surface curvature. Lower potential energy for the electron corresponds to higher electric 
potential, hence electric field lines come out of that region and flow into the region of lower electric potential. Thus, electrons 
in the superconductor, a macroscopic quantum system, keep their total energy
constant throughout the superconductor by adjusting their potential $and$ kinetic energy,  giving rise to a macroscopically
inhomogeneous charge distribution, in contrast to a classical system or normal metal that
has  a macroscopically uniform charge distribution that gives lowest macroscopic potential energy.

In Fig. 4 we show an example of the electric field arising for a body with an egg-like geometry, composed of half an oblate and half a 
prolate ellipsoid fused together. Here the electric field distribution is more complicated, however it can be seen that it follows the general pattern
discussed in the previous paragraph, with electric field lines coming out of regions of lower curvature and coming into regions of higher curvature.
It is interesting to note that we find numerically that the electric field configuration for samples of this shape has no net electric dipole moment, which would be allowed
by symmetry in the sample shown in Fig. 4 (not in the sample of Fig. 3). We have not been able to find an analytic explanation of this finding.

These electric fields around superconductors should be experimentally detectable in small ($\mu$-size) superconducting samples. High quality   smooth surfaces are 
required to prevent trapping of charge in localized states and imperfections and to allow the electric field lines to extend away from the surface. The magnitude of the expected electric fields near the surface is of order 
thousands of Volts/cm\cite{ellipsoid}, and these shape-dependent electric fields should only exist  when the sample is 
superconducting. `Conventional' superconductors such as $Pb$ and $Nb$ would be the best systems to check this prediction\cite{electrospin}.
Indirect evidence that these electric fields exist is seen in the Tao effect\cite{tao1,tao2}.

    \begin{figure}
 \resizebox{7.5cm}{!}{\includegraphics[width=9cm]{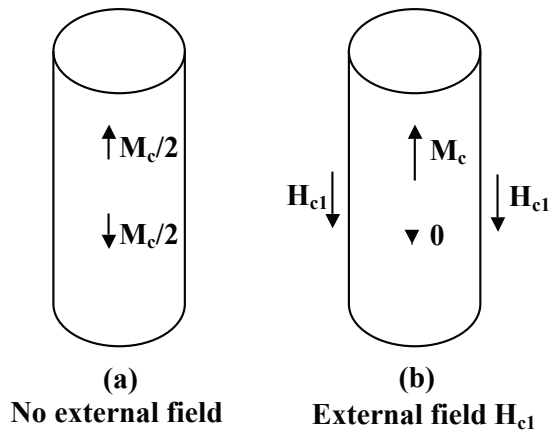}}
 \caption {Corrected Figure 1, taking into account the ground state orbital magnetic moments predicted by this theory.
 }
 \label{figure2}
 \end{figure}
 
Associated with the inhomogeneous charge distribution our theory predicts 
that electrons move in orbits of radius $2\lambda_L$ and  carry orbital angular momentum $\hbar/2$ in
direction opposite to the spin angular momentum\cite{sm}. From this results   a macroscopic spin current flowing within a London penetration depth of the 
surface of superconductors\cite{electrospin}. The speed of the carriers is given by
\beq
\bold{v}_\sigma^0=-\frac{\hbar}{4m_e\lambda_L}\hat{\sigma}\times \hat{n}      
\eeq
with $\hat{n}$ the outward pointing normal to the surface. The current within $\lambda_L$ of the surface for electrons of spin $\sigma$ is
\beq
J_\sigma=n_\sigma e v_\sigma^0=\frac{n_\sigma}{2}\mu_B
\eeq
witn $n_\sigma=n_s/2$, and the resulting magnetization is
\beq
M_\sigma=\frac{J_\sigma \lambda_L}{c}=\frac{n_\sigma}{2}\mu_B=\frac{M_c}{2}
\eeq
pointing in opposite direction to the magnetization due to the electron spin, which is twice as large because of the gyromagnetic factor $g=2$. Thus, 
the situation depicted in Fig. 1 gets modified to what is shown in Fig. 5. The initial magnetization for spin $\sigma$, $M_c/2$, results from the 
compensation of spin magnetization $M_c$ and orbital magnetization $M_c/2$ in opposite directions. The external field $H_{c1}$ generates
magnetization $M_c$ in one direction, and in the process brings the orbital motion of one of the spin components to a stop\cite{sm}, at which point the
system goes normal.

The Meissner effect in this theory is explained as resulting from the orbital expansion and associated charge expulsion in the transition to
superconductivity\cite{kinetic3}. A distinctive property of superconductors within this theory  is that they exhibit macroscopic
quantum zero-point motion, and the same is proposed to be true in   superfluid $^4He$\cite{helium}.

\end{document}